\documentclass[12pt,epsf]{article}  
\usepackage{epsf,color,latexsym}
\newcommand{\Ss}{{\cal S}}

\newcommand{\wt}{\widetilde}
\newcommand{\tphi}{\wt\phi}

\newcommand{\tSigma}{{\widetilde\Sigma}}
\newcommand{\hSigma}{{\widehat\Sigma}}
\newcommand{\poly}{{\rm PolyLog}}

\newcommand{\be}{\begin{equation}}
\newcommand{\ee}{\end{equation}}
\newcommand{\ben}{\begin{eqnarray}\displaystyle}
\newcommand{\een}{\end{eqnarray}}
\newcommand{\refb}[1]{(\ref{#1})}

\begin{document}

{}~ \hfill\vbox{\hbox{hep-th/0105312}
\hbox{UUITP-02/01}}\break 

\vskip 3.0cm

\centerline{\large \bf  Quantum Corrections in}
\vspace*{4.0ex}

\centerline{\large \bf  $p$-adic String Theory}
\vspace*{1.5ex}

\vspace*{8.0ex}

\centerline{\large \rm Joseph A. Minahan\footnote{E-mail: joseph.minahan@teorfys.uu.se}}
\vspace*{2.5ex}
\centerline{\large \it Department of Theoretical Physics}
\centerline{\large \it Box 803, SE-751 08 Uppsala, Sweden}
\vspace*{3.0ex}

\vspace*{2.5ex}

\vspace*{4.5ex}
\medskip
\centerline {\bf Abstract}

\bigskip
We compute  loop corrections  in $p$-adic open string field theory. 
We argue that quantum effects induce a
pole with $m^2\sim - \ln g$  for the open string
field at the locally stable vacuum.  We
also compute the one loop effective potential and show that the potential
develops an imaginary piece
 when the  field becomes tachyonic.
\vfill \eject
\baselineskip=17pt

%\sectiono{Introduction}

The $p$-adic string turns out to be a useful model for testing
Sen's conjectures \cite{senconj} on tachyon condensation in
open string field theory \cite{BFOW,0003278,0102071}.  
The classical solutions of $p$-adic string field theory can be explicitly
found. They have constant descent relations \cite{BFOW,0003278}
and a fluctuation spectrum that matches $p$-adic string amplitudes
on D-brane backgrounds \cite{0003278,0102071}.  The tachyon potential
in $p$-adic string field theory also contains a classically stable vacuum
where the mass of the tachyon field is pushed up to infinity \cite{BFOW,
0003278}, further justifying the Sen conjectures.

Of course, the Sen conjectures have been intensely studied in conventional
bosonic string field theory
\cite{cubiccalc} and in the superstring analog \cite{superchecks},
where the results have been nothing short of impressive.
More recently, it now looks possible to find an exact verification
of the Sen conjectures by expanding the string fields
about the closed string vacuum 
\cite{0012251,0102112,0105058,0105059,0105168}.

However, the similarity between $p$-adic string theory and ordinary
string field theory goes beyond the verification of the Sen conjectures.
For example,
$p$-adic string field theory in  the $p\to1$ limit 
\cite{0009103} reduces to 
the two derivative truncation of the tachyon effective action
\cite{0008231} in boundary string
field theory \cite{9210065,0009103,0009148,0009191}, 
suggesting that results for $p$-adic
strings are applicable to boundary string field theory.
Perhaps an even more compelling reason for studying $p$-adic string
theory is the surprising similarity between the $p$-adic D-branes
and those recently found in vacuum cubic string field theory,
where the string action is expanded about the closed string vacuum
\cite{0102112,0105058,0105059,0105168}.  In both cases
the D-branes have gaussian profiles.  Furthermore, the $p$-adic string does not suffer
from strong coupling problems at the local minimum, a featured shared
with  cubic string field theory.

Therefore, the success of the $p$-adic models in qualitatively verifying
the Sen conjectures and the further similarity to ordinary string
theory suggests that they might be useful for exploring
other qualitative features of string field theory.   In this paper
we will consider quantum corrections in $p$-adic string field theory. 
The calculations are significantly easier than in a conventional scalar
field theory due to the nature of the kinetic term.  Loop calculations turn
out to be gaussian integrals.

The loop calculations lead to  some interesting results.  
In particular, we will compute
the self energy correction about the stable vacuum, where we find that
the classical  infinite mass pole is pulled down to a finite value.  We
also show that the analytic continuation of the 
one loop effective action develops an imaginary
part if the tachyon field
is large enough.  This occurs before the open string
vacuum is reached. The presence of the imaginary piece should
not come as a great surprise.  This just reflects the instability
of  the open string vacuum --- the one-loop partition function
for a tachyon scalar field formally has an IR divergence which can be continued
to a finite, but imaginary piece.  

One might speculate whether or not these results carry over qualitatively to 
convential string field theory.  If so then the conclusion would be that
ordinary bosonic string theory has massive  states in the closed
string vacuum
 which are the remnants of
open strings from the classical D-brane solutions.  Perhaps one lesson
that is applicable to vacuum cubic string field theory is that the quantum
corrections about the closed string vacuum are surprisingly easy to compute.
The relative simplicity of the BRST operator and the 
classical solutions
about the closed string vacuum suggest that quantum corrections will be
simpler than what one usually finds for string loop corrections.  

The tachyon effective action
for the $p$-adic string was derived in \cite{BFOW} and is given by
\begin{equation}\label{effact}
\Ss=-\frac{1}{g^2}\frac{p^2}{p-1}\int d^Dx
\left[\frac{1}{2}\phi p^{-\frac{1}{2}\Box}\phi-
\frac{1}{p+1}\phi^{p+1}\right].
\end{equation}
The action was derived assuming that $p$ is prime, but once this action is
in place, there does not appear to be any compelling reason to keep $p$ a
prime number.  In fact, it was shown in \cite{0009103} that the limit $p\to1$ yields
the two derivative truncation for  the effective  action in boundary
string field theory.  

The tree-level  open string vacuum is at $\phi=1$, while the locally
stable solution is at
$\phi=0$.   The pole position as a function of the background field $\phi$ is
\be\label{polecl}
q^2=2+\frac{2(p-1)}{\ln p}\ln \phi.
\ee
Thus the solution at $\phi=1$ has a tachyon pole, while the
solution at $\phi=0$ has  its pole pushed to  infinite $m^2$.  
Hence, in this vacuum
the open string states disappear from the spectrum.

However,  eq. \refb{polecl} would
seem to indicate  that once quantum fluctuations are included, a pole with a finite $m^2$ is possible.  Since the
mass is infinite only when $\phi=0$,  fluctuations about the $\phi=0$ solution
could allow a state to propagate.  We will explicitly verify this.

We will compute the quantum corrections
 by expanding about the $\phi=0$ solution.  As previously stated
there is no strong coupling problem at this vacuum, so the results are
trustworthy, at least at weak string coupling.
For the first part of
 this analysis we will assume that $p$
is an odd integer  If $p$ is odd, then the potential is
an even function of $\phi$ and the closed string vacuum remains at $\phi=0$
after quantum corrections are included.  Later, when we compute the one-loop
effective action, we will assume that $p$ is arbitrary.  If $p$ is not
an odd integer, then the effective action will have a one-point function,
shifting the value of $\phi$ at the minimum.  In this case we will see that
there is still a pole at finite mass squared.

The Feynman rules are easily derived from \refb{effact}.  The propagator
with Euclidean momentum $k$ is $p^{-\frac{1}{2}k^2}$, while the $p+1$ point
vertex is given by 
\begin{equation}
p!g^{p-1}\left(\frac{p-1}{p^2}\right)^{\frac{p-1}{2}}.
\end{equation}
  The first graph we consider is
the two point diagram in Figure 1, whose existence requires that $p$ be
an odd integer.  This is the lowest order correction
to the two-point function and has $\frac{p-1}{2}$ loops.
The calculation of this graph is straightforward and is given by
\begin{equation}
-\Sigma=p!g^{p-1}\left(\frac{p-1}{p^2}\right)^{\frac{p-1}{2}}\frac{2^{\frac{p-1}{2}}}{\left(\frac{p-1}{2}\right)!}\prod_{i=1}^{\frac{p-1}{2}}\int\frac{d^Dk_i}{(2\pi)^D} p^{-\frac{1}{2}k_i^2}
\end{equation}
where $\frac{2^{\frac{p-1}{2}}}{\left(\frac{p-1}{2}\right)!}$ is
a symmetry factor.
Hence we find that
\begin{equation}
-\Sigma=\frac{p!}{\left(\frac{p-1}{2}\right)!}\left(\frac{2g^2(p-1)}{p^2(2\pi\ln p)^{\frac{D}{2}}}\right)^{\frac{p-1}{2}}.
\end{equation}

\begin{figure}[!ht]
\leavevmode
\begin{center}
\epsfbox{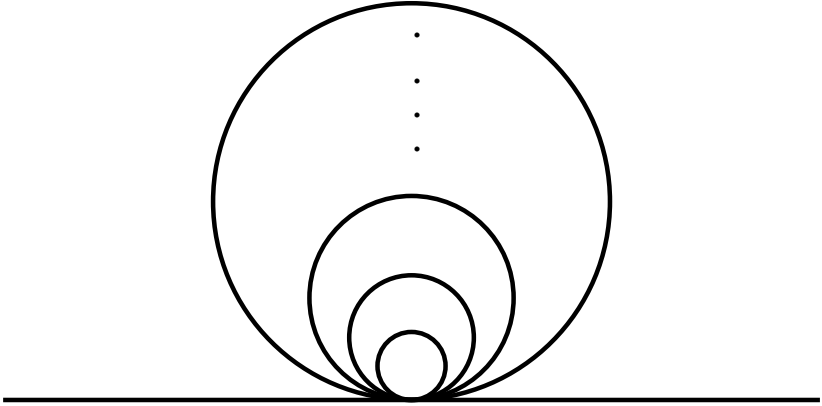}
\end{center}
\caption[]{\small Lowest order correction to the two point function, with
$\frac{p+1}{2}$ loops.}  \label{f1}
\end{figure}

Including this term in the effective action, we see that there is a pole
when
\begin{equation}
p^{\frac{1}{2}q^2}+\Sigma=0.
\end{equation}
Hence, we find a pole at
\begin{equation}\label{mass1}
q^2=\frac{p-1}{\ln p}\ln\left(\frac{2g^2(p-1)}{p^2(2\pi\ln p)^{\frac{D}{2}}}\right)+\frac{2}{\ln p}\ln\left(\frac{p!}{\left(\frac{p-1}{2}\right)!}\right).
\end{equation}
In the weak coupling limit when $g<<1$, $q^2\sim \ln g$, so the pole is
for a propagating field with a large, but finite mass squared.

We can check the robustness of this result by computing the next contribution
to the two point function.  The relevant diagram is shown in figure 2.
\begin{figure}[!ht]
\leavevmode
\begin{center}
\epsfbox{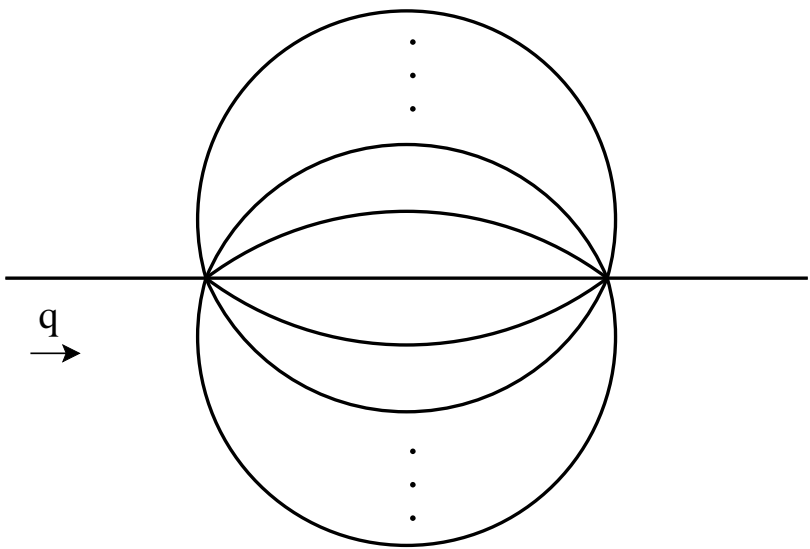}
\end{center}
\caption[]{\small Next order correction to the two point function, with
$p-1$ loops.}  \label{f2}
\end{figure}
This diagram has $p-1$ loops, and depends on the external momentum
$q$.  Using the Feynman rules and including symmetry factors, we find
that the contribution $\tSigma(q)$ is
\be\label{mass2}
-\tSigma(q)=(p!)^2g^{2(p-1)}\left(\frac{p-1}{p^2}\right)^{p-1}\frac{1}{p!}
\int\prod_i^{p}\frac{d^Dk_i}{(2\pi)^D}p^{-\frac{1}{2}k_i^2}(2\pi)^D\delta^D(q+\sum k_i)
\ee
Explicitly doing one of the integrals to get rid of the $\delta$-function
and then completing the squares in the gaussian integrals, 
we find that \refb{mass2} reduces to
\be\label{mass2i}
-\tSigma(q)=p!p^{-\frac{D}{2}}\left(\frac{g^2(p-1)}{p^2(2\pi\ln p)^{\frac{D}{2}}}\right)^{p-1}p^{-\frac{1}{2p}q^2}.
\ee
Substituting the solution to $q$ in \refb{mass1}, we see that $\tSigma$
is of the order
\be
\tSigma\sim -g^{(p-1)(2-\frac{1}{p})}.
\ee
Hence, the contribution from $\tSigma(q)$ is suppressed by an
extra factor of order $g^{(p-1)^2/p}$ as compared to $\Sigma$.  Thus the
solution for the pole in \refb{mass1} is valid at weak coupling as long
as $p\ne 1$.

We now turn to the one loop effective action.  The relevant diagrams are
shown in figure 3.
\begin{figure}[!ht]
\leavevmode
\begin{center}
\epsfbox{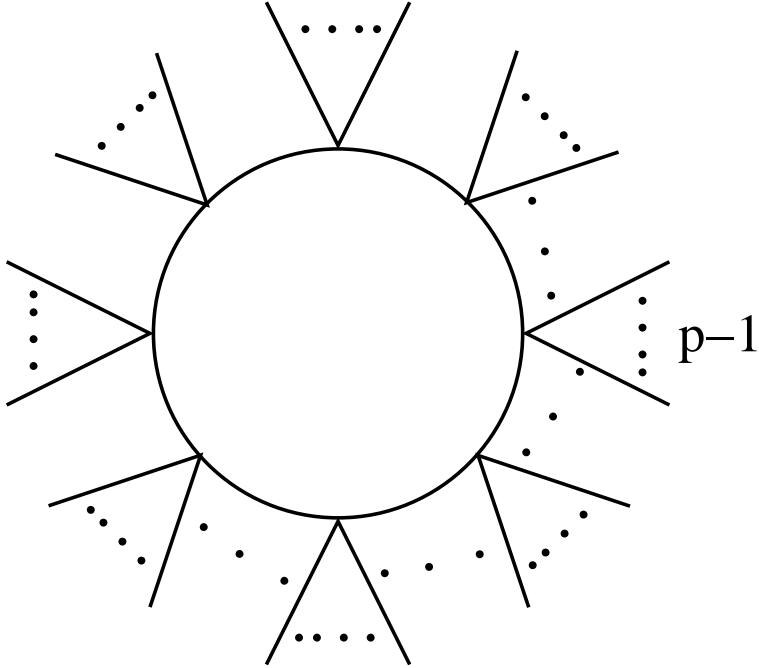}
\end{center}
\caption[]{\small Contributions to the one loop effective action.}  \label{f3}
\end{figure}
Clearly, there are one loop contributions of the form $\phi^{n(p-1)}$. Expanding the potential to second order in the fluctuations $\tphi$, we have
\be
V(\phi)=\frac{1}{p+1}\phi^{p+1}+\phi^p\tphi+\frac{p}{2}\phi^{p-1}\tphi^2+...
\ee
Therefore, taking into account the symmetry factors, the one loop correction to the effective potential is
\begin{eqnarray}\label{1loop}
V_{1loop}(\phi)&=&-\sum_{N=1}\frac{1}{2N}(p\phi^{p-1})^N\int \frac{d^Dk}{(2\pi)^D}
p^{-\frac{1}{2}Nk^2}\nonumber\\
&=& -\frac{1}{2(2\pi \ln p)^{\frac{D}{2}}}\sum_{N=1} \frac{1}{N^{1+\frac{D}{2}}}
(p\phi^{p-1})^N\nonumber\\
&=&-\frac{1}{2(2\pi \ln p)^{\frac{D}{2}}}\poly\left(1+\frac{D}{2},p\phi^{p-1}\right),
\end{eqnarray}

The series in \refb{1loop} formally diverges if $p\phi^{p-1}>1$.  However, the
result can be analytically continued, yielding a finite potential, but one with
an imaginary component.  The real part is combined with the tree level potential, showing that for weak coupling, $\phi$ has a small shift at the open string vacuum.  Since the open string vacuum is still close to $\phi=1$, we see that
the potential here has an imaginary part.  This of course arises from the
tachyon mode  in the open string vacuum.  In particular, note that the imaginary
piece develops when the value of $q^2$ in \refb{polecl} changes sign. 
In string theory, the contribution of the tachyon to
the one loop partition function, after analytic
continuation, is finite but has an imaginary part.

Let us now consider the case where $p=2$.  We might expect this case to
most closely mirror cubic string field theory, since the tree level
action in both cases is cubic.  The action is no longer
invariant under $\phi\to -\phi$, so the quantum corrections will induce a one-point function and hence a
shift in the local minimum.  We can easily compute this shift using \refb{1loop}.  Adding \refb{1loop} to \refb{effact}, we see that the new vacuum $\phi_0$
approximately satisfies
\be
\phi_0\approx \frac{1}{4(2\pi\ln2)^{\frac{D}{2}}}g^2.
\ee

However, the contribution to $\tSigma(q)$ in \refb{mass2i} is also a one-loop
contribution when $p=2$, and because of the $q^2$ dependence in $\tSigma(q)$,
the $\tSigma(q)$ term will dominate in the computation for the pole.  To see
this explicitly, note that the new equation for the pole is
\be\label{mass3}
2^{\frac{1}{2}q^2}-\phi_0-\frac{g^2}{2(4\pi\ln 2)^{\frac{D}{2}}}\ 2^{-\frac{1}{4}q^2}=0.
\ee
There is no corresponding term $\Sigma$ as in
\refb{mass1} because $p-1$ is odd.  Clearly, the last term on the lhs of \refb{mass3} will dominate
over the $\phi_0$ term and the pole is approximately at
\be\label{mass3i}
q^2=\frac{4}{3\ln2}\left(\ln g^2-\ln2-\frac{D}{2}\ln(4\pi\ln2)\right)
\ee
Thus, we find the same sort of $\ln g$ dependence for the pole position.

We should check that   terms that contribute to next order
 are suppressed when evaluated at the pole.
For $p=2$, one such term has the form in
figure 4.
\begin{figure}[!ht]
\leavevmode
\begin{center}
\epsfbox{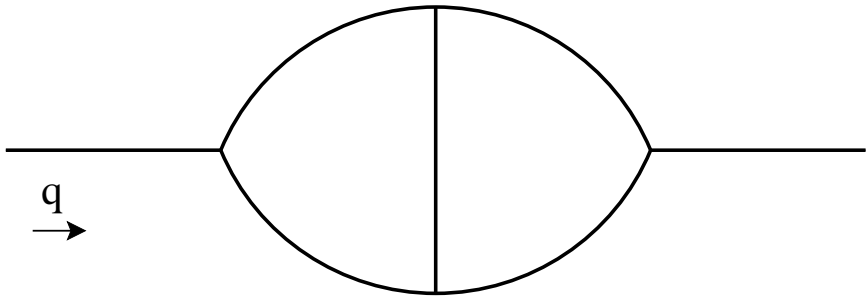}
\end{center}
\caption[]{\small Two loop contribution to the $p=2$ two-point function.}  \label{f4}
\end{figure}
Including the symmetry factors, we find that the contribution of this graph
is
\begin{eqnarray}
-\hSigma(q)&=&g^4\int \frac{d^Dk_1}{(2\pi)^D}\frac{d^Dk_2}{(2\pi)^D}
2^{-\frac{1}{2}(k_1^2+k_2^2+(k_1-k_2)^2+(k_1-q)^2+(k_2-q)^2)}\nonumber\\
&=&\frac{g^4}{(4\pi\sqrt{2})^D}\ 2^{-\frac{3}{4}q^2}.
\end{eqnarray}
Hence, this term is suppressed for the value of $q^2$ in \refb{mass3i}.

Of course we would like to carry out the analagous computations in cubic
string
field theory.\footnote{There has been recent work on loop corrections for 
boundary
string field theory \cite{0104099,0104164,0105098,0105227,0105238}.}
Here we will just speculate on the results we might find.

First, we should expect to find finite results, just as in the
$p$-adic case.  
Ordinarily there are two types of divergences in one-loop
string partition functions.  The first is due to the open string
tachyon, which we 
have previously argued can be analytically 
continued to a finite contribution.  
In the closed string vacuum there should not be
any imaginary contribution since the tachyon is absent.  
The second type of divergence comes from the closed string zero
momentum poles.  Part of this divergence comes from the closed string tachyon,
which can be analytically continued to a finite contribution.
The other divergences are due to the graviton and dilaton, which are 
removed through the Fischler-Susskind mechanism.  In any case, all of
the closed string divergences are proportional to the coupling of the
closed string fields to the D-brane which are in turn proportional
to the tension of the brane.  In the closed string vacuum, the branes are
absent, so there is nothing for the closed string tachyon, 
graviton and dilaton to couple to.  Hence
the closed string  divergences should not be present in the closed string
vacuum.

Second, since quantum corrections encode fluctuations about the closed
string vacuum,
it would seem that the mechanism is in place to produce
  poles from the open string fields
in the closed string vacuum, perhaps  with the same sort of $\ln g$ behavior. 
In the tree level closed string vacuum, all states with finite poles
are BRST trivial.  If there are to be nontrivial states, then quantum 
corrections must modify the form of the BRST operator.

There have
been many recent works  about how closed strings can appear
in open string theories 
\cite{9901159,0002223,0009061,0010240,0011009,0011226,0012081}.  One might
guess  that the pole described here represents a closed string state.  
In ordinary
string theory, the closed string spectrum is independent of $g$ at 
weak coupling, so a pole with $\ln g$ dependence seems strange.  On the other
hand, in $p$-adic string theory one does not have the usual closed string
states; for instance there are no infinities in the one-loop diagrams corresponding to massless closed string or off-shell tachyon exchange.  So perhaps
the state with $\ln g$ behavior is the unique $p$-adic closed string state.

In cubic string field theory,   it is likely that if one pole appears
then infinitely many will appear.  Hence, it seems possible to construct
an infinite tower of finite mass poles.  However, it still seems as if
the tower would have the masses pushed to infinity as $g$ approaches zero,
making the identification of these poles with closed string states problematic.
Hopefully  calculations can be carried out soon to shed some light on
this problem.

\bigskip
\noindent {\bf Acknowledgments}:
I thank Barton Zwiebach for  helpful discussions and for comments on
the manuscript. 
I also thank  the CTP at MIT for 
hospitality during the course of this work.
This work was
supported in part by the NFR.

\end{document}